\documentclass[%
 reprint,
superscriptaddress,
longbibliography,
 amsmath,amssymb,
 aps,
 prb,
]{revtex4-1}

\usepackage{sidecap}
\sidecaptionvpos{figure}{c}
 \usepackage[T1]{fontenc}
 
\usepackage{graphicx}
\usepackage{dcolumn}
\usepackage{bm}
\usepackage[
colorlinks=true,
linkcolor=blue,
citecolor=blue,
filecolor=blue,
urlcolor=blue]{hyperref}

\usepackage{amsmath,amsfonts,
	amsgen,amsbsy,amsopn,amstext,amssymb
}
\usepackage{epstopdf}
\usepackage{xcolor}

\usepackage{color,soul} 

\renewcommand{\i}{i}

\definecolor{redish}{RGB}{255, 173, 231}
\definecolor{bluish}{RGB}{205, 224, 247}

\begin{document}

\preprint{}

\title{Stealth and equiluminous materials for scattering cancellation and wave diffusion}
\author{S. Kuznetsova}\email{svetlana.kuznetsova@univ-lemans.fr}
\affiliation{Laboratoire d'Acoustique de l'Universit\'e du Mans (LAUM), UMR CNRS 6613, Institut d'Acoustique - Graduate School (IA-GS), CNRS, Le Mans Universit\'e, France}

\author{J.-P. Groby}
\affiliation{Laboratoire d'Acoustique de l'Universit\'e du Mans, LAUM - UMR 6613 CNRS, Le Mans Universit\'e, Avenue Olivier Messiaen, 72085 LE MANS CEDEX 9, France}

\author{L.M. Garcia-Raffi}
\affiliation{Instituto de Matem\'atica Pura y Applicada (IUMPA), Universitat Polit\`ecnica de Val\`encia, Camino de vera s/n, 46022, Valencia, Spain.}

\author{V. Romero-Garc\'ia}\email{vicente.romero@univ-lemans.fr}
\affiliation{Laboratoire d'Acoustique de l'Universit\'e du Mans, LAUM - UMR 6613 CNRS, Le Mans Universit\'e, Avenue Olivier Messiaen, 72085 LE MANS CEDEX 9, France}

\date{\today}

\begin{abstract}
  We report a procedure to design 2-dimensional acoustic structures with prescribed scattering properties. The structures are designed from targeted properties in the reciprocal space so that their structure factors, i.e., their scattering patterns under the Born approximation, exactly follow the desired scattering properties for a set of wavelengths. The structures are made of a distribution of rigid circular cross-sectional cylinders embedded in air. We demonstrate the efficiency of the procedure by designing 2-dimensional stealth acoustic materials with broadband backscattering suppression independent of the angle of incidence and equiluminous acoustic materials exhibiting broadband scattering of equal intensity also independent of the angle of incidence. The scattering intensities are described in terms of both single and multiple scattering formalisms, showing excellent agreement with each other, thus validating the scattering properties of each material.
\end{abstract}

\pacs{Valid PACS appear here}
\maketitle

\section{Introduction}
Scattering of waves by a many-body system is an interdisciplinary topic of interest in several branches of science and technology ranging from statistical mechanics or condensed matter to wave physics. When such a system is excited by an incident wave, the incoming energy is both scattered and absorbed by the obstacle. This results in a scattering pattern that is highly dependent on the geometry and size of the scatterer distribution as well as on the frequency-dependent properties of the material of the constituent scatterers. 
The manipulation of wave scattering has long been a topic of discussion in various classical areas of physics including electromagnetism \cite{Electromagn}, photonics \cite{Optics} and acoustics \cite{Acoustics}, but in recent decades significant attention has been paid to artificial structured media to control waves. Photonic\cite{PhotC1,PhotC2,PhotC3} or phononic\cite{PhonC1,PhonC2,PhonC3} crystals, hyperuniform and stealth materials\cite{uche2004constraints,torquato2002random,batten2008classical,Torquato16,Torquato15} as well as metamaterials \cite{engheta2006metamaterials, Liu00, Fang06, wong2017optical} are just a few examples of many-body systems to control the scattering of the incident wave. 

Ordered structures, such as photonic \cite{PhotC1,PhotC2,PhotC3} and phononic\cite{PhonC1,PhonC2,PhonC3, Groby08} crystals, exhibit multiple overlapping Bragg diffraction peaks and thus peculiar dispersion relations that can serve as efficient tools for the control of wave scattering. Metamaterials are complex structures that can be tuned and reconfigured to control the scattering of the incident wave through the resonance of their constituent building blocks \cite{PhonC3, Craster13, Romero19}. Another way of manipulating wave scattering is offered by disordered structures, in which the phase transition between the wave diffusion and localization regimes occurs due to the interference of the waves scattered in the media \cite{wiersma1997localization,hu2008localization}. Among the disordered systems, stealth materials are characterized by the stealthiness, i.e. the suppression of the single scattering of the incident radiation for a given subset of wave vectors\cite{torquato2002random, batten2008classical}. Recently, one dimensional stealth acoustic materials have been numerically and experimentally designed to provide stealthiness on demand robust to losses\cite{romero2019stealth}. A subclass of stealth materials is given by the stealth hyperunifom materials for which transparency appears in a subset of wave vectors around the origin\cite{uche2004constraints,torquato2002random,batten2008classical,Torquato16,Torquato15}. The relevance of hyperuniformity appeared in condensed matter physics when classical systems of particles interacting with certain soft long-ranged pair potentials could counterintuitively freeze into hyperuniform states. In other words, these systems were counter to the common expectation that liquids freeze into crystal structures with high symmetry. An increasing interest was focused on stealth hyperuniform materials, or simply on hyperuniform materials, as they have been used to design networks with complete band gaps comparable in size to those of a photonic/phononic crystal, while at the same time maintain statistical isotropy, enabling waveguide geometries not possible with photonic/phononic crystals as well as high-density disordered transparent materials. \cite{gkantzounis2017hyperuniform,florescu2009designer,man2013isotropic,man2013photonic,Band_gap, batten2008classical}. Another important class of disordered many-body systems are equiluminous materials\cite{batten2008classical}, which scatter waves uniformly in all directions. Such omnidirectional diffusion could play an important role in improving room acoustics by avoiding unwanted reflections \cite{cox2009acoustic,Acoustics,PhysRevX.7.021034}.

Materials with targeted scattering properties are usually designed by inverse methods, i.e., their structure parameters are extracted from the scattering data. Although this approach relies on an ill-posed problem \cite{tikhonov1977solutions,colton2019inverse}, various material design tools based on targeting the scattering properties of the structure have been implemented in both wave physics and condensed matter. Inverse approach \cite{Inverse,Inverse4,Inverse5} consists in optimizing the inter-particle interactions (thus minimizing some energetic characteristics) leading to self-assembling from a simpler condition. Optimization methods operating in direct space rely on zero-temperature and near-melting temperature technique to obtain lattice ground state configurations \cite{Inverse,Inverse1,Inverse1_Erratum,Inverse2,Inverse2_Erratum,Inverse3} and collective-coordinates technique for soft matter and disordered ground states \cite{Collective,Inverse_Hyper}. Usual numerical methods include black-box optimization benchmarking \cite{Black_box}, probabilistic \cite{Cohn9570,C3SM27785B} and genetic algorithms \cite{Genetic1,Genetic2} to name a few. A flat acoustic lens \cite{Hakansson,Hakansson2} focusing sound at a predefined point, a photonic-crystal-based structure\cite{Hakansson3} performing requested optical tasks, or a sonic demultiplexing device \cite{Hakansson1} spatially separating several wavelengths were designed using a genetic algorithm in conjunction with the multiple scattering theory (MST) \cite{MTS, Schwan} to optimize a cluster of scatterers. A 2-dimensional low loss acoustic cloak for air-born sound has also been designed by means of genetic algorithm and simulated annealing \cite{Garcia-Chocano}. Nonlinear conjugate gradient algorithm has been used to optimize a graded porous medium composed of a periodic arrangement of ordered unit cells to provide the optimal acoustic reflection and transmission \cite{Vicent1}. Recently scattering suppression of electromagnetic waves for prescribed wavelengths and directions has been achieved by pre-assigning the scattering properties in the reciprocal space and using generalized Hilbert transform \cite{PhysRevA.98.013822}.


In this work, we design disordered 2-dimensional (2D) acoustic structures consisting of rigid circular cross-sectional cylinders embedded in air. These structures are designed to present prescribed scattering properties when excited by a plane wave. We target the information on the scattering pattern in the reciprocal space and use an optimization procedure, which optimizes the positions of scatterers to ensure the targeted scattering properties. A weak scattering approach is followed, which allows us to characterize the system by its structure factor. This factor turns out to be proportional to the scattered intensity and only depends on the scatterer positions when they are identical. Therefore, the optimization procedure finds the distribution of scatterers producing the targeted structure factor values by fixing the scattering properties in the reciprocal space and as a consequence the desired scattering properties. The polar scattering pattern of the optimized distribution of scatterers is first evaluated from the representation of the structure factor in the reciprocal space by using the von Laue formulation. This scattering pattern is then evaluated independently by the MST, which is a self-consistent method accounting for all orders of scattering. Comparison of the results of the two methods allows us to validate the approximation of weak scattering and consequently the results. We apply the proposed approach to design and describe 2D stealth and equiluminous materials showing broadband back-scattering suppression and broadband equally intense scattering respectively, independently of the angle of incidence.
 



\section{Scattering in many-body systems: Structure factor and multiple scattering theory}
\label{sec:SG}

\begin{figure*}
\includegraphics[width=18cm]{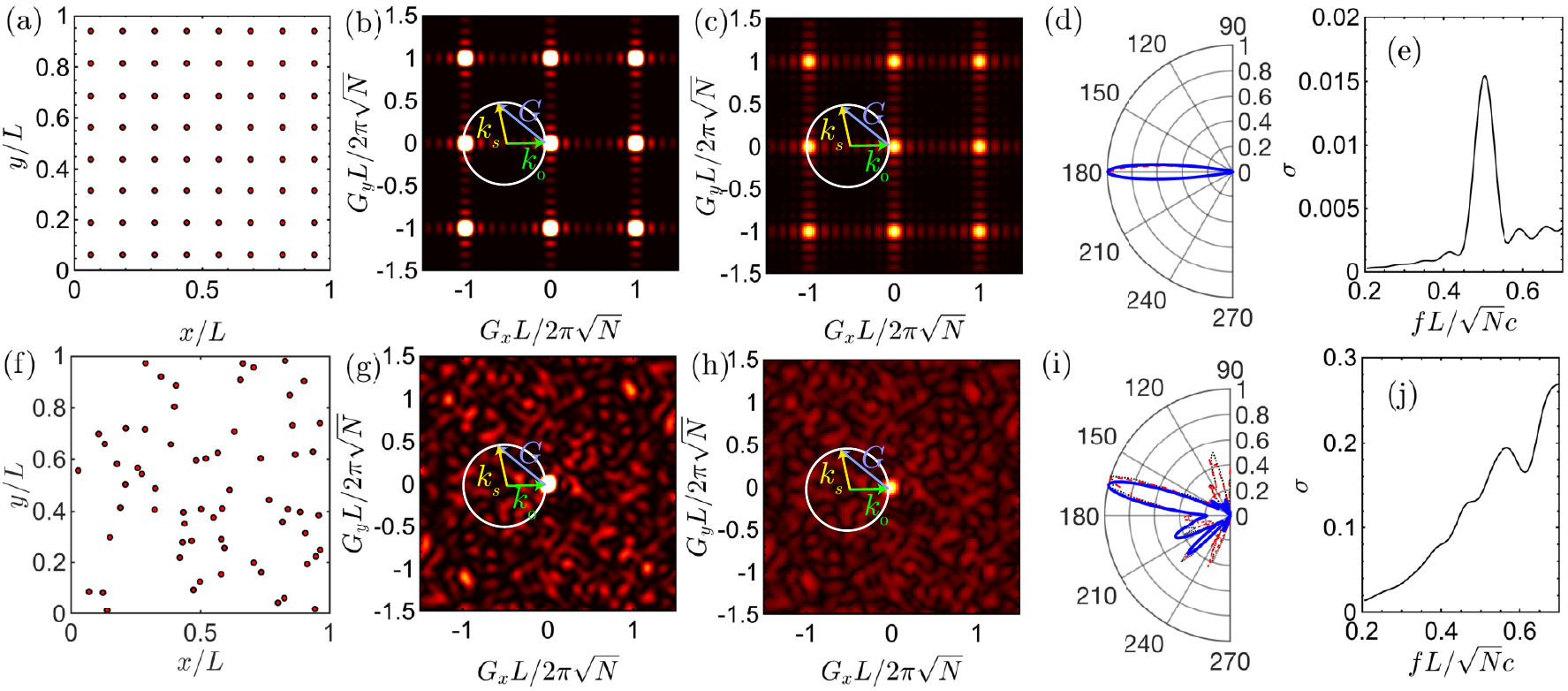}
\caption{(Color online) Scattering by an array of $N$ scatterers radiated by an incident plane wave characterized by the wavevector $\vec{k}_0$. (a) and (f) represent a periodic and a  random distribution of $N=64$ cylinders with $R_0=L/100$ respectively.\footnote{The coordinates of these distributions of points are provided in the Supplementary Material} (b) and  (g) show the representation of the structure factor $S(\vec{G})$ for the periodic and the random distribution respectively. (c) and (h) show the representation of the spatial Fourier transform $|\mathcal{FT}(\vec{G})|$ for the periodic and the random distribution respectively. In (b-c) and (g-h) the incident wavevector and the scattered wavevector are related through the Ewald circumference with the vectors of the reciprocal space. (d) and (i) show the polar distribution of the normalized scattered far field intensity between $\theta=[90, 270]$ degrees for the periodic and the random distributions respectively. Black dotted (red dashed line) [continuous blue line] represents the results obtained from the $S(\vec{G})$ ($|\mathcal{FT}(\vec{G})|^2$) [$|P^f_s(\theta, \omega)|^2$]. (e) and (j) represents the scattering cross section of the periodic and random distributions respectively.}
\label{fig:fig1}
\end{figure*}

We are interested in the scattering of acoustic waves by structures made of a distribution of $N$ rigid cylindrical scatterers with circular cross-section of identical radius $R_i=R_0$ and located at positions $\vec{r}_i$ with $i=1,..., N$. These $N$ scatterers are embedded in a square area $\Omega$ of the direct space of side $L$.  
We assume weak scattering, i.e., the amplitude of the scattered field is small compared to that of the incident field. Under this condition, we assume that the Born approximation is satisfied. Strictly speaking, Born approximation corresponds to the case in which the incident field to the $i$-th cylinder is only composed of the incident wave, i.e. no scattererd waves by the other scatterers impinges the $i$-th scatterer. For the geometries considered in this work, the weak scattering approximation is thus valid for low filling fractions and when the scatterer radii are small compared to the wavelength (see Appendix \ref{app:MST} for more details).

This discrete system can be characterized by the following scalar function defined in the spatial (direct) domain $\Omega$ as
\begin{eqnarray}
\rho(\vec{r})=f(\vec{r})\ast\sum_{i=1}^N\delta(\vec{r}-\vec{r}_i),
\end{eqnarray}
where $\ast$ is the convolution operator, $\delta(\vec{x})$ is the Dirac's delta and $f(\vec{r})$ is the transparency of the scatterer, defined without loss of generality as
\begin{eqnarray}
f(\vec{r})=\left\{
\begin{array}{ll}
      0 & \textrm{if}\;|\vec{r}|>R_0,\\
      1 & \textrm{if}\;|\vec{r}|\leq R_0.
\end{array} 
\right. 
\end{eqnarray}
Under these assumptions, the amplitude of the scattered wave is proportional to the spatial Fourier transform of $\rho(\vec{r})$, $\mathcal{FT}(\vec{G})$, where $\vec{G}$ is a vector of the reciprocal space. This follows from the well known theory in optics that the diffraction pattern of a structure is equal to the product of the diffraction pattern of the base element and that of the array\cite{Taillet}. Through this work we assume a time harmonic dependence of the type $e^{-\imath \omega t}$ where $\omega$ the angular frequency. With this, we simply end with
\begin{eqnarray}
\mathcal{FT}(\vec{G})=f(\vec{G})\times \sum_{i=1}^{N} e^{-\imath \vec{G} \vec{r}_{i}}.
\end{eqnarray}
Therefore, the scattered intensity is given by
\begin{eqnarray}
I(\vec{G})= |f(\vec{G})|^2 \times \sum_{i=1}^{N}\sum_{j=1}^{N}e^{-\imath\vec{G} (\vec{r}_{i}-\vec{r}_{j})},
\end{eqnarray}
where $f(\vec{G})$ is known as the atomic structure factor and only depends on the geometry of the scatterer as our scatterers are considered rigid. Thus, the scattered intensity can be simply written as
\begin{eqnarray}
I(\vec{G})=|f(\vec{G})|^2NS(\vec{G}),
\label{eq:I}
\end{eqnarray}
where 
\begin{eqnarray}
S(\vec{G})= {\frac {1}{N}}\sum_{i=1}^{N}\sum_{j=1}^{N}e^{-\imath\vec{G} (\vec{r}_{i}-\vec{r}_{j})}={\frac {1}{N}}\left|\sum _{j=1}^{N}{e} ^{\imath \vec{G} \vec{r} _{j}}\right|^2,
\label{eq:SG}
\end{eqnarray}
is the structure factor. It should be noted here that the structure factor only depends on the position of the scatterers in $\Omega$. Moreover, we notice that $S(\vec{G})=S(-\vec{G})$ (see Appendix \ref{app:SG} for more details and additional demonstrations). The structure factor is extensively used in condensed matter or wave physics to describe the scattering of an incident wave by a given structure made of a distribution of scatterers. However, multiple scattering effects are neglected, although this approach has the benefit of allowing fast predictions.

The wave scattering by a distribution of scatterers can effectively be more precisely described by the MST \cite{MTS, Schwan}. The far-field expression of the scattered field provided by the MST (see Appendix \ref{app:MST} for more details) when the structure is radiated by a plane wave with wave vector $\vec{k}_0$, is given by
\begin{eqnarray}
p_{s}^{f}=P^f_s(\theta,\omega)\sqrt{\frac{k}{\imath 2\pi r}}e^{\imath k r},\;\;r\rightarrow\infty,
\end{eqnarray}
where $k=|\vec{k}_0|$ and the far-field scattered amplitude $P^f_s(\theta,\omega)$, at angle $\theta$, reads as
\begin{eqnarray}
P^f_s(\theta,\omega)=\frac{2}{k} \sum_{i=1}^N e^{-\imath k |\vec{r}_i| \cos{(\theta-\theta_{\vec{r}_i})}} \sum_{n}(-i)^nA_n^ie^{\imath n\theta},\nonumber\\
\end{eqnarray}
with $\theta_{\vec{r}_i}$, the azimuthal angle of the position vector of the $i$-th cylinder $\vec{r}_i$ and $A_n^i$, the scattering coefficients of the $i$-th cylinder calculated by MST. The scattered far-field intensity is thus proportional to $I(\theta, \omega)\propto |P^f_s(\theta,\omega)|^2$. The scattering cross section of the scatterer distribution when excited by a plane wave $e^{\imath k x}$ is also computed by applying the Optical Theorem (see  Appendix \ref{app:MST}) via
\begin{eqnarray}
\sigma=-2\textrm{Re}(P^f_s(0,\omega))=-\frac{4}{k}\textrm{Re}\left( \sum_{i=1}^N e^{-\imath k x_{\vec{r}_i}} \sum_{n}(-i)^nA_n^i\right).\nonumber\\\label{eq:sigma}
\end{eqnarray}

In order to illustrate the different results provided by each method and the interpretation of the scattering in the reciprocal space, the scattering properties of both a periodic and a random distribution of $N=64$ rigid cylinders of radii $R_0=L/100$ are analyzed (array of $8\times 8$ rigid cylinders in the periodic case). The random distribution has been generated by choosing random positions of the scatterers and avoiding overlapping between them. The periodic [random] distribution of scatterers is plotted in Fig. \ref{fig:fig1}(a) [\ref{fig:fig1}(f)].\footnote{The coordinates of these distributions of points are provided in the Supplementary Material}. Figures \ref{fig:fig1}(b-c) [\ref{fig:fig1}(g-h)] respectively depict $S(\vec{G})$ and $|\mathcal{FT}(\vec{G})|$ in the reciprocal space. In the periodic case, both $S(\vec{G})$ and $\mathcal{FT}(\vec{G})$ consist of a periodic pattern of sinc-type functions. The maxima appear due to the periodicity of the square distribution at $\vec{G}=\left(n\frac{2\pi\sqrt{N}}{L},m\frac{2\pi\sqrt{N}}{L}\right)$ with $(n,m) \in \mathbb{Z}^2$. In the random case, the representations in the reciprocal space are not periodic as shown in Fig. \ref{fig:fig1}(g-h). In both periodic and random cases, a hot spot in the center that represents the forward scattering is exhibited. The parity of both $S(\vec{G})=S(-\vec{G})$ and $|\mathcal{FT}(\vec{G})|=|\mathcal{FT}(-\vec{G})$| is clearly visible.

To interpret the scattering produced by these distributions, we first discuss how the scattering is directly interpreted in the reciprocal space using the von Laue formulation. Let us consider that the system is excited by an incident plane wave the wavevector of which is $\vec{k}_0=k(\vec{e}_x,0)$, with $\vec{e}_x$ the unitary vector along the $x$ direction and $k=|\vec{k}_0|$. We choose $k=\pi N/L$ in this particular example to analyze the Bragg scattering in the periodic case. This wavevector $\vec{k}_0$ is represented in Figs. \ref{fig:fig1}(b-c) [Figs. \ref{fig:fig1}(g-h)] for the periodic [random] case pointing one of the points of the reciprocal space. The von Laue formulation of the wave diffraction\cite{Ashcroft1976} stipulates that the difference between the vector of the scattered wave, $\vec{k}_s$, and that of the incident wave, must be a vector of the reciprocal space, i.e., $\vec{k}_s-\vec{k}_0=\vec{G}$, for constructive interference to occur. Assuming elastic scattering, $|\vec {k} _{s}|\equiv k_s=|\vec {k_{0}} |=k=2\pi /\lambda$, only the vectors pointing non zero values of $S(\vec{G})$ along the Ewald sphere\cite{Ashcroft1976} can lead to scattered waves for 3D problems. This sphere of radius $k_0$ is centered at the origin of $\vec{k}_0$ in the reciprocal space. More precisely, all the possible scattered waves are given by the Ewald sphere. The scattered wavevectors, $\vec{k}_s$, are then given by the vector connecting the center of the sphere and the points along this sphere having a non null value of $S(\vec{G})$. 
The scattering is finally activated along the direction given by these vectors $\vec{k}_s$. This discussion is valid for any dimension, the Ewald sphere is a circumference in 2D of radius $k_0$ centred in $\vec{k}_0$, and is given by the limits of a segment of length $2k_0$ centred in $\vec{k}_0$ in 1D.

Following this procedure, the values of $S(\vec{G})$ and $|\mathcal{FT}(\vec{G})|$ along the Ewald circumference between $\theta~=~[90, 270]$ degrees can be evaluated and, this polar distribution provides the back scattering produced by systems. Figure \ref{fig:fig1}(d)  [Figure \ref{fig:fig1}(i)] shows the polar pattern of the scattered field by the periodic [random] distribution. Both $S(\vec{G})$ and $|\mathcal{FT}(\vec{G})|$ present a strong back scattering around 180$^o$ as expected from the Bragg scattering. We then compare the results with the scattered far field as calculated by the MST, i.e., when accounting for all the scattering orders. We conclude that the Bragg scattering is produced and that the results are very close to those given by both $S(\vec{G})$ and $|\mathcal{FT}(\vec{G})$|. For the random scatterer distribution, scattering along more directions than in the periodic case is expected, because more vectors are possible in the reciprocal space. Some directions predicted by both $S(\vec{G})$ and $|\mathcal{FT}(\vec{G})|$ are nevertheless  missing when comparing the results with the scattered far field as calculated with the MST. This is due to the fact that scatterers are too close to each other in some area of the distribution for the weak scattering approximation to be valid  as shown in Fig. \ref{fig:fig1}(f). In that case, the most realistic modeling is that provided by the MST. However, it should be noted that the main directions of scattering are captured by both $S(\vec{G})$ and $\mathcal{FT}(\vec{G})$.

To conclude this analysis concerning the scattering by a periodic and a random pattern of scatterers, Figs. \ref{fig:fig1}(e) and \ref{fig:fig1}(j) depict the scattering cross section as calculated with the MST for the periodic and the random cases respectively. The Bragg interference produces a peak of scattering at the Bragg frequency $f_{Bragg}=Nc/2L$, where $c$ is the speed of sound in the host fluid material, for the periodic distribution while the scattering cross section almost continuously increases with frequency for the random distribution.


\section{Material design tool}
Depending on the values of the structure factor in the reciprocal space, different kinds of materials can be designed\cite{batten2008classical}. In general, the following constraint can be imposed on the structure factor
\begin{eqnarray}
S(\vec{G})=S_0\;\;\;\textrm{for}\;\;\;K_1\leq|\vec{G}|\leq K_2,
\end{eqnarray}
where $K_1$ and $K_2$ are respectively the lower and the upper limits of an area of the reciprocal space for which $S(\vec{G})$ has a constant value $S_0$.  Stealth materials present zero structure factor, $S_0=0$, for a general set of reciprocal vectors. Hyperuniform materials\cite{Torquato02, Torquato03, batten2008classical, florescu2009designer, man2013isotropic, man2013photonic, Torquato15, Torquato16, gkantzounis2017hyperuniform} are specific type of stealth materials for which $K_1=0$, i.e., infinite-wavelength density fluctuations vanish up to $K_2=K$. $d$-dimensional hyperuniform materials are in addition characterized by $\chi=\frac{1}{\pi^{d-1}}\left(\frac{KL}{2N}\right)^d$, which represents the relative fraction of constrained degrees of freedom for a fixed reciprocal-space exclusion-sphere of radius $K$. Finally, equiluminous materials present a structure factor that is other than zero, i.e., $S_0\neq 0$ for specified wavevectors in the reciprocal space.

\begin{figure*}
\includegraphics[width=18cm]{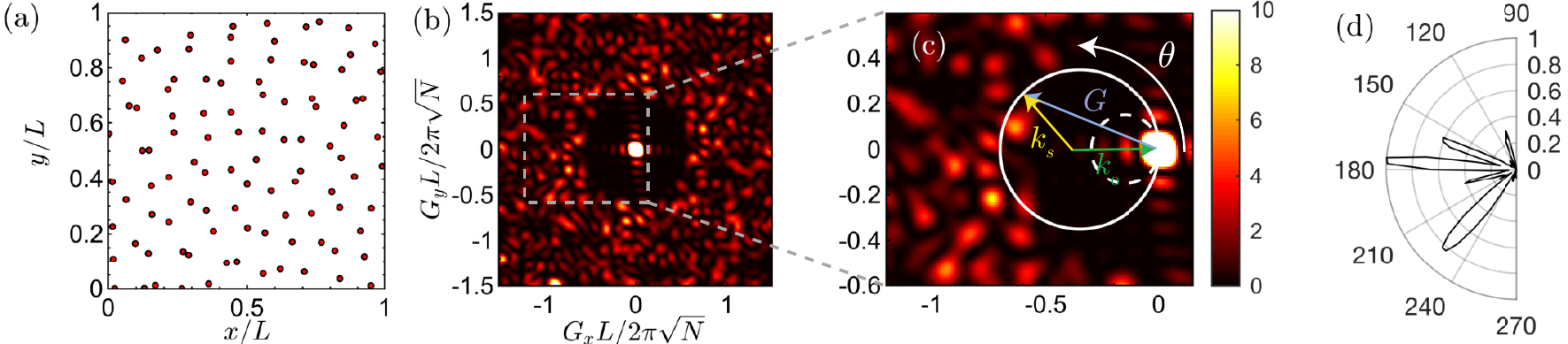}
\caption{(Color online) Structure factor and scattering produced by a Hyperuniform material made of $N=100$ cylinders with $R_0=L/100$. (a) represents the point distribution for the Hyperuniform material (with $\chi=0.196$).\footnote{The coordinates of this distribution of points are provided in the Supplementary Material} (b) shows the structure factor. (c) is a zoom the region of interest in (b). We plot the Ewald circumference corresponding to an incident wavevector $\vec{k}_0$. (d) represents the polar plot of the normalized scattered intensity calculated from the structure factor for the Hyperuniform (at $k_0=k_s=0.35\frac{2\pi\sqrt{N}}{L}$ [white circumference in (c)] between $\theta=[90, 270]$ degrees in order to analyze the back scattering components.}
\label{fig:fig2}
\end{figure*}


In this section, we present a methodology to design structured materials with prescribed scattering features. In contrast to real-space methods, the desired scattering characteristics are introduced directly in the reciprocal space via the structure factor and an optimization procedure is used to find, the scatter distributions that gives rise to the targeted value of structure factor for a set of wavelengths. The cost function that is minimized during the optimization procedure is a function of the structure factor itself $S(\vec{G})$ [Eq. (\ref{eq:SG})]. For a given limit of wavevectors, the structure factor must have a target value $S_{0}$ for all the wavevectors in domain $\Lambda$. The objective function reads as
\begin{eqnarray}
\phi(\vec{r}_1,\dots,\vec{r}_N)=\sum_{\vec{G}\in\Lambda}\left(S(\vec{G})-S_{0}\right),
\label{eq:potential}
\end{eqnarray}
and is subjected to the following constrains to avoid overlapping of scatterers of radius $R_0$,
\begin{eqnarray}
|\vec{r}_i-\vec{r}_j|\geq 2R_0\;\;\;\forall i\neq j.
\end{eqnarray}
We note that Eq.~(\ref{eq:potential}) is already norm $L_2$ as $S(\vec{G})$ is already norm $L_2$. The optimization algorithm looks for distribution of scatterers $\vec{r}_i$ that minimizes the Eq. (\ref{eq:potential}). Stealth, Hyperuniform, and Equiluminous materials or more generally any kind of materials with targeted properties in the reciprocal space can be designed.

As an example, Figs. \ref{fig:fig2}(a-c) show a hyperuniform material made of a distributions of $N=100$ rigid cylinders with $R_0=L/100$ with $\chi=0.196$ designed by the present procedure.\footnote{The coordinates for these distribution of points can be found in the supplementary material.} Figure \ref{fig:fig2}(b) represents the structure factor of the corresponding scatterer distribution. The scattering suppression area with $S(\vec{G})=0$ between $K_1=0$ and $K_2=0.5\frac{2\pi\sqrt{N}}{L}$ is clearly visible. Following the discussion on the interpretation of the scattering in the reciprocal space given in Section \ref{sec:SG}, Fig. \ref{fig:fig2}(c) shows the Ewald circumferences for two different incident wavevectors $\vec{k}_0=(0.35\frac{2\pi\sqrt{N}}{L}\vec{e}_x,0)$ (white continuous line) and $\vec{k}_0=(0.15\frac{2\pi\sqrt{N}}{L}\vec{e}_x,0)$ (white dashed line). These two situations correspond respectively to cases where the Ewald circumference is either partially or completely within the scattering suppression area where $S(\vec{G})=0$. In both cases, the points along the circumference falling in the region $S(\vec{G})=0$ do not produce scattering. For this reason, hyperuniform materials suppress the scattering of incident radiation at low frequencies. This is the case of the Ewald circumference for $\vec{k}_0=(0.15\frac{2\pi\sqrt{N}}{L}\vec{e}_x,0)$ [white dashed line in Fig. \ref{fig:fig2}(c)]. In the opposite, strong back scattering occurs for $\vec{k}_0=(0.35\frac{2\pi\sqrt{N}}{L}\vec{e}_x,0)$. Figure \ref{fig:fig2}(d) shows the corresponding polar diagrams of the normalized scattered intensities between $\theta=[90, 270]$ degrees as calculated with Eq.~(\ref{eq:I}). The intensity is normalized with respect to its maximum value for $\vec{k}_0=(0.35\frac{2\pi\sqrt{N}}{L}\vec{e}_x,0)$. Therefore, back scattering is clearly exhibited when $\vec{k}_0=(0.35\frac{2\pi\sqrt{N}}{L}\vec{e}_x,0)$  while no back scattering occurs when $\vec{k}_0=(0.15\frac{2\pi\sqrt{N}}{L}\vec{e}_x,0)$. Note that the asymmetry of the polar diagram in Fig. \ref{fig:fig2}(d) around the direction $\theta=180$ is a direct consequence of the disorder of the scatterer distribution.

\begin{figure*}
\includegraphics[width=18cm]{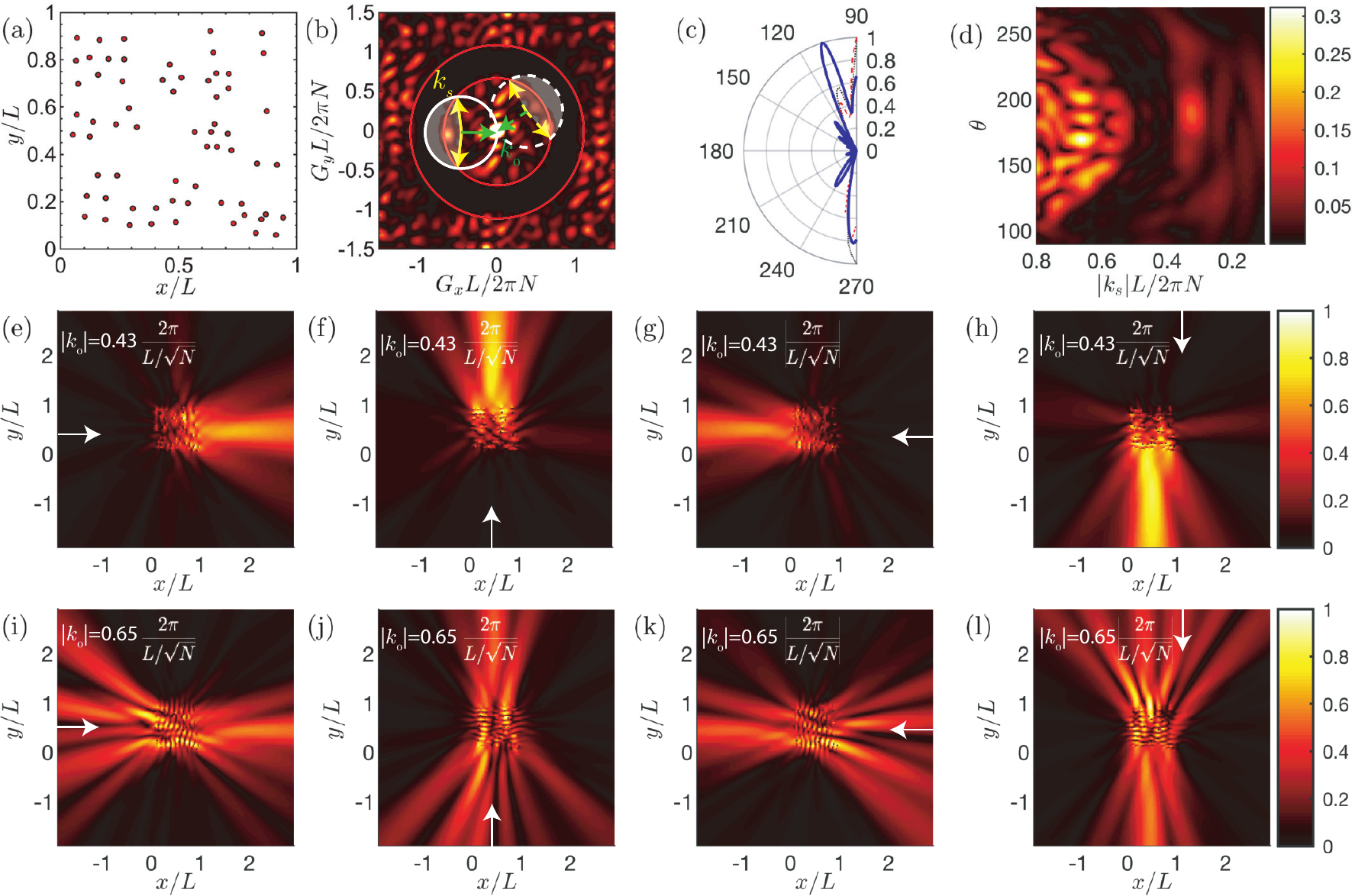}
\caption{(Color online) Scattering properties of an stealth material for broadband and omnidirectional back scattering suppression. (a) Stealth distribution.\footnote{The coordinates of this distribution of points are provided in the Supplementary Material} (b) Structure factor $S(\vec{G})$ of the stealth material. Ewald circumference distribution for two different incident directions of the same wave. Whitish area shows the scattering suppression for these two examples.  Red circles in (b) represent the limits of scattering suppression, $K_1=0.75\frac{2\pi\sqrt{N}}{L}$ and $K_2=1.1\frac{2\pi\sqrt{N}}{L}$. Green (yellow) arrow represents the incident (scattered) plane wave. (c) Polar plot of the scattered field, $|p_s^f(\theta,\omega)|$ produced by this Stealth material obtained from the structure factor (dotted black line), the Fourier transform (dashed red line) and the MST (continuous blue line) for $k=0.43\frac{2\pi\sqrt{N}}{L}$. (d) $\theta-|k|$ map of the scattered pressure field $|p_s^f(\theta,\omega)|$ obtained from MST for an incident wavevector $\vec{k}_0=(0.43\frac{2\pi\sqrt{N}}{L}\vec{e}_x,0)$. (e-h) Scattered pressure field $|p_s|$ for an incident wave with $|\vec{k}_0|=0.43\frac{2\pi\sqrt{N}}{L}$ (inside the scattering suppression region) along 0$^o$, 270$^o$, 180$^o$, 90$^o$ of incidence respectively. (i-l) Normalized scattered pressure field $|p_s|$ for an incident wave with $|\vec{k}_0|=0.65\frac{2\pi\sqrt{N}}{L}$ (outside the scattering suppression region) along 0$^o$, 270$^o$, 180$^o$, 90$^o$ of incidence respectively.}
\label{fig:fig3}
\end{figure*}

\section{Results}
In this section we show the results for different materials with targeted scattering properties using the structure factor. We compare the results obtained from the structure factor, the spatial Fourier transform, and the MST.

\begin{figure*}
\includegraphics[width=18cm]{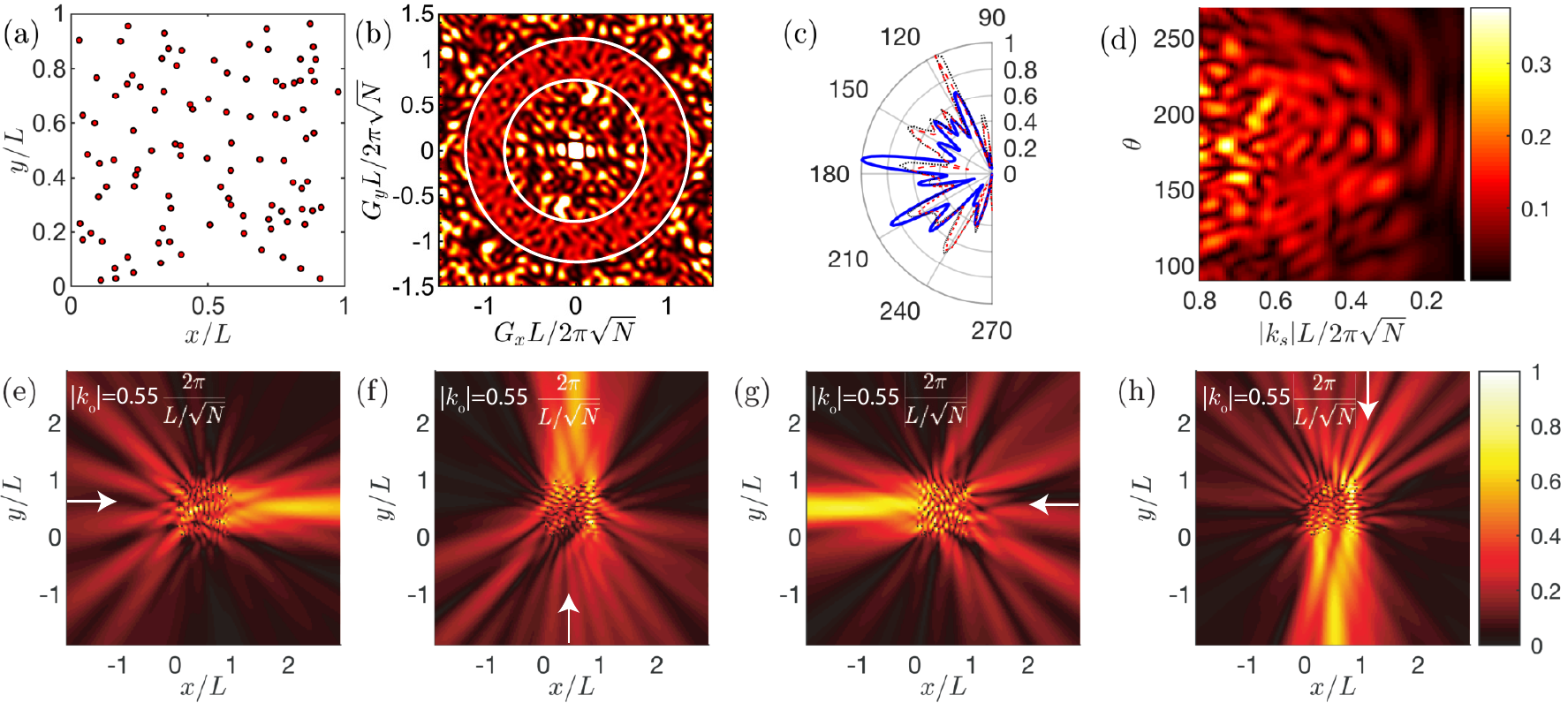}
\caption{(Color online) Scattering properties of an equiluminus material for broadband and omnidirectional diffusion. (a) Equiluminus distribution.\footnote{The coordinates of this distribution of points are provided in the Supplementary Material} (b) Structure factor $S(\vec{G})$ of the Equiluminus material. White circles in (b) represent the limits of equally intense scattering area, $K_1=0.8\frac{2\pi\sqrt{N}}{L}$ and $K_2=1.2\frac{2\pi\sqrt{N}}{L}$ (with $S_0=1$). (c) Polar plot of the scattered field, $|p_s^f(\theta,\omega)|$ produced by this Equiluminus material obtained from the structure factor (dotted black line), the Fourier transform (dashed red line) and the MST (continuous blue line) for $k=0.55\frac{2\pi\sqrt{N}}{L}$. (d) $\theta-|k|$ map of the scattered pressure field $|p_s^(\theta, \omega)|$ obtained from MST for an incident wavevector $\vec{k}_0=(0.55\frac{2\pi\sqrt{N}}{L}\vec{e}_x,0)$. (e-h) Scattered pressure field $|p_s|$ for an incident wave with $|\vec{k}_0|=0.55\frac{2\pi\sqrt{N}}{L}$ (inside the equally intense scattering region) along 0$^o$, 270$^o$, 180$^o$, 90$^o$ of incidence respectively.}
\label{fig:fig4}
\end{figure*}

\subsection{Broadband back-scattering suppression independent of the angle of incidence}
We start by analyzing the properties of a stealth material presenting a scattering suppression area between $K_1=0.75\frac{2\pi\sqrt{N}}{L}$ and $K_2=1.1\frac{2\pi\sqrt{N}}{L}$. A distribution of $N=64$ identical cylindrical rigid scatterers with $R_0=L/100$ is considered.  Figure \ref{fig:fig3}(a) shows the scatterer distribution that minimizes Eq. (\ref{eq:potential}) resulting from the material design tool. Figure \ref{fig:fig3}(b) depicts the corresponding structure factor showing the scattering suppression area between the imposed limits. Note that this particular shape of scattering suppression area implies that the back scattering is completely suppressed for any frequency and for any incident wavevector in this area. In Fig. \ref{fig:fig3}(b), two different Ewald circumferences are shown for two different incident directions at the same frequency. The whitish areas represent those of the suppressed scattered wavevectors at this particular frequency. It should be noted here that fixing the frequency of an incident wave having scattered wavevectors in the suppression zone, the suppressed scattering remains the same for any incident direction. 

Figure \ref{fig:fig3}(c) shows the polar plot of the scattered field $|p_s^f(\theta,\omega)|$ by this stealth material as calculated with the structure factor (dotted black line), the Fourier transform (dashed red line) and the MST (continuous blue line) for $\vec{k}_0=(0.43\frac{2\pi\sqrt{N}}{L},0)$. The back scattering is almost suppressed as evidenced by the polar plot. Figure \ref{fig:fig3}(d) represents the $\theta-|k|$ map of $|p_s^f(\theta,\omega)|$ obtained from the MST. The scattering suppression is clearly in good agreement with the results presented Fig. \ref{fig:fig3}(b). This evidences the broadband back scattering suppression independent of the incident angle. 

Figures \ref{fig:fig3}(e-h) finally show the scattered pressure field $|p_s|$ for an incident plane wave with $k_0=0.43\frac{2\pi\sqrt{N}}{L}$ for four different incident directions, $\vec{k}_0=(k_0,0)$, $\vec{k}_0=(0,k_0)$, $\vec{k}_0=(-k_0,0)$, and $\vec{k}_0=(0,-k_0)$ respectively (white arrow). The back scattering components are strongly reduced for each angle of incidence.

As predicted with the Ewald circumference in the structure factor map, the forward component is the most important. Although the values of the structure factor, i.e., the scattered intensity, inside the suppression area are not exactly zero, the independence of the back scattering with respect to the incidence angle is still remarkable. In order to prove the strong effect of this back scattering suppression, we show the scattered pressure distribution for a frequency at which the scattered wavevectors fall outside the scattering suppression region. Figures \ref{fig:fig3}(i-l) show the scattered pressure field for $k_0=0.65\frac{2\pi\sqrt{N}}{L}$ at the same incident directions. Both the back scattering and the forward scattering are of equal importance.



\subsection{Broadband equally intense scattering independently of the angle of incidence}

Contrary to the stealth materials, we design an equiluminous materials thanks to the proposed material design tool that produce broadband equally intense scattering independent of the angle of incidence. In this section, we discuss an equiluminous material of equally intense scattering area between $K_1=0.8\frac{2\pi\sqrt{N}}{L}$ and $K_2=1.2\frac{2\pi\sqrt{N}}{L}$. For this particular case, and without loss of generality, we consider a distribution of $N=100$ rigid scatterers with $R_0=L/100$. Figure \ref{fig:fig4}(a) shows the scatterer distribution minimizing Eq. (\ref{eq:potential}) with $S_0=1$. Figure \ref{fig:fig4}(b) shows the corresponding structure factor showing the equally intense scattering area between the imposed limits [white circles in Fig. \ref{fig:fig4}(a)]. Similarly to the stealth material, this particular shape of equally intense scattering area implies that the back scattering is equally distributed for any frequency and any incident direction (see Fig. \ref{fig:fig3}(b) for the Ewald circumference representation) in this area. Nevertheless, the values of $S(\vec{G})$ inside the equally intense scattering are not completely homogeneous due to the discrete character of the proposed design contrary those of the stealth material. This finds translation in a quasi-equally intense scattering pattern. The values of $S(\vec{G})$ are however clearly smother inside the target area than outside.

Figure \ref{fig:fig4}(c) depicts the polar plot of the scattered field $|p_s^f(\theta,\omega)|$ by this equiluminus material as calculated with the structure factor (dotted black line), the Fourier transform (dashed red line) and the MST (continuous blue line) for $k=0.55\frac{2\pi\sqrt{N}}{L}$. Contrary to the stealth material, the back scattering is almost evenly distributed along the angles and quasi-equally intense as evidenced by the polar plot. Figure \ref{fig:fig4}(d) represents the $\theta-|k|$ map of $|p_s^f(\theta,\omega)|$ obtained from the MST. The quasi-equally intense scattering is in good agreement with the results plotted in Fig. \ref{fig:fig4}(b). This evidences the broadband back scattering behaviour of the structure independent of the incident angle. 

Figures \ref{fig:fig4}(e-h) show the scattered pressure field $|p_s|$ for an incident wave with $k_0=0.55\frac{2\pi\sqrt{N}}{L}$ for four different incident directions, $\vec{k}_0=(k_0,0)$, $\vec{k}_0=(0,k_0)$, $\vec{k}_0=(-k_0,0)$, and $\vec{k}_0=(0,-k_0)$ respectively. The back scattering components are angularly distributed with quasi-equal intensity for each angle of incidence. Although the values of the structure factor, i.e., the scattered intensity, inside the equally intense scattering area are not exactly constant, the quasi-equal intense back scattering independent of the angle of incidence is still remarkable.

\section{Conclusions}

Heterogeneous materials formed by a set of scatterers embedded in a host material with tailored properties are a useful tool for the control and manipulation of acoustic, electromagnetic and matter waves. In this work, we present a methodology based on prescribing the scattering properties of the system in the reciprocal space, i.e., prescribing its structure factor, to later obtain the spacial distribution of scatterers with the corresponding scattering properties.
The developed methodology was applied to construct stealth and equiluminous materials. The scattered intensity was first obtained from the structure factor based on their proportionality in the weak scattering approximation. The results were validated using the multiple scattering approach that accounts for all the scattering orders. The scattered intensity patterns obtained by these two approaches are in excellent agreement having similar angular distributions. First we have designed a stealth system that exhibits broadband back-scattering suppression independently of the incidence directions, having zero structure factor in the given frequency range and as a consequence, a close to zero scattered intensity. Second, we have designed an equiluminous system that provides broadband diffusion independently of the incident direction, having non zero constant structure factor in the desired range of frequencies. Although the scattered intensity is not exactly the same at different scattering angles (we have a discrete distribution of scatterers), the scattering pattern is still quasi-intense and is smoother inside the target frequency range than outside it. The proposed methodology has proved itself as a powerful tool to design and characterize disordered many-body systems with preassigned scattering properties.



\appendix

\section{Multiple Scattering Theory}
\label{app:MST}
We consider that the $N$ cylinders of radius $R_i$ are located at $\vec{r}_i$ with $i=1,...,N$ to form the distribution in the $x-y$ plane. The system is excited by a plane wave of the form $p_0(\vec{r})=e^{\imath kx}$ with a temporal dependence of the type $e^{-\imath\omega t}$. The scattered wave by the $i$-cylinder can be written as
\begin{eqnarray}
p_s(\vec{r},\vec{r}_i)=\sum_nA_n^i {\text{H}}_n (k|\vec{r}-\vec{r}_i|)e^{\imath n\theta_{(\vec{r}-\vec{r}_i)}},
\end{eqnarray}
where $\text{H}_n$ is the $n$-th order Hankel function of first type.
The total field incident to $i$-th cylinder $p_{in}^i(\vec{r})$ is a superposition of the direct contribution from the incident wave $p_0(\vec{r})$ and the scattererd waves from all the other scatterers
\begin{eqnarray}
p_{in}^i(\vec{r})=p_0(\vec{r})+\sum_{j=1,j\neq i}^N p_s(\vec{r},\vec{r}_j).
\label{eq:born1}
\end{eqnarray}
This incident wave on the $i$-th cylinder can be expressed as follows
\begin{eqnarray}
p_{in}^i(\vec{r})=\sum_nB^i_n{\text{J}}_n(k|\vec{r}-\vec{r}_i|)e^{\imath n \theta_{\vec{r}-\vec{r}_i}},
\end{eqnarray}
where $\text{J}_n$ is the $n$-th order Bessel function of first type.
We now express the scattered field by the $i$-th cylinder in the vicinity of the $j$-th cylinder. To do so, we use the Graff's theorem:
\begin{eqnarray}
p_s(\vec{r},\vec{r}_j)=\sum_n C_n^{j,i}{\text{J}}_n(k|\vec{r}-\vec{r}_i|)e^{\imath n \theta_{\vec{r}-\vec{r}_i}},\\~\forall |\vec{r}-\vec{r}_i| \in [R^j, |\vec{r}_j-\vec{r}_i|-R_i[,
\end{eqnarray}
with 
\begin{eqnarray}
C_n^{j,i}=\sum_l A_l^j {\text{H}}_{l-n} (k|\vec{r}_i-\vec{r}_j|)e^{\imath (l-n)\theta_{\vec{r}_i-\vec{r}_j}}.
\end{eqnarray}

The incident plane wave is then represented in the $i$-th cylinder coordinate system, via
\begin{eqnarray}
p_0(\vec{r})&=&e^{\imath k x}=e^{\imath k x_j}e^{\imath k |\vec{r}-\vec{r_j}|\cos{(\theta_{\vec{r}-\vec{r}_j})}}.
\end{eqnarray}
At this stage, we use the Jacobi-Anger expansion to expand the term $e^{\imath k |\vec{r}-\vec{r_j}|\cos{(\theta_{\vec{r}-\vec{r}_j})}}$ upon Bessel functions:
\begin{eqnarray}
e^{\imath k |\vec{r}-\vec{r_j}|\cos{(\theta_{\vec{r}-\vec{r}_j})}}=\sum_n\imath^n{\text{J}}_n(k|\vec{r}-\vec{r}_j|)e^{\imath \theta_{\vec{r}-\vec{r}_j}}.
\end{eqnarray}
Therefore, we end with
\begin{eqnarray}
p_0(\vec{r})&=&\sum_n S_n^i {\text{J}}_n(k|\vec{r}-\vec{r}_j|)e^{\imath \theta_{\vec{r}-\vec{r}_j}},
\end{eqnarray}
where
\begin{eqnarray}
S_n^i =\imath^ne^{\imath k x_j}.
\end{eqnarray}
The factor $e^{\imath k x_j}$ plays the role of a complex amplitude which depends on the horizontal projection of the position of the $j$-th scatterer, $x_j$.


Now that we have expressed all the acoustic fields involved in the problem in the vicinity of the $i$-th cylinder, we can obtain the following system of equations:
\begin{eqnarray}
B_n^i=S_n^i+\sum_{j=1,j\neq i}^N \sum_lA_l^j{\text{H}}_{l-n}(k|\vec{r}_i-\vec{r}_j|)e^{\imath (l-n)\theta_{\vec{r}_i-\vec{r}_j}}.\nonumber\\
\end{eqnarray}
At this stage, the $S_n$ are known, but both $B_n$ and $A_l$ are unknown. The rigid boundary condition provides another equation
relating them. At the interface of the $i$-th cylinder, we have
\begin{eqnarray}
\left. \frac{1}{\rho_0}\frac{\partial p_{ext}}{\partial r}\right|_{r=R_i}&=0,
\end{eqnarray}
giving rise to
\begin{eqnarray}
B_n^i&=&-\frac{\textrm{H}_n'(kR_i)}{\textrm{J}_n'(kR_i)}A_n^i\equiv \Gamma_n^i A_n^i,
\end{eqnarray}
where the primes represent derivative. Therefore the amplitudes of the scattered and the incident fields on the $i$-th cylinder can be related by means of $\Gamma_n^i$.

Finally the system of equations can be written as follows,
\begin{eqnarray}
\Gamma_n^i A_n^i-\sum_{j=1,j\neq i}^N \sum_l G_{n,l}^{j,i} A_l^j=S_n^i,
\label{eq:sys}
\end{eqnarray}
where
 \begin{eqnarray}
 G_{n,l}^{j,i}={\text{H}}_{l-n}(k|\vec{r}_i-\vec{r}_j|)e^{\imath (l-n)\theta_{\vec{r}_i-\vec{r}_j}}\;\;\textrm{for}\;\;i\neq j.\;\;\;
 \end{eqnarray}
 This system of equations is solved for every frequency by truncating the infinite sums. A good estimation for this truncation is 
 \begin{eqnarray}
l=n=\textrm{floor}\left(kR_{max}+4.05(kR_{max})^{1/3}\right)+10,\;\;\;\;\;\;
 \end{eqnarray}
with $R_{max}=\textrm{max}(R_i)$. Once the system is solved, the coefficients $A_n^i$ are known and the total pressure can be obtained from
 \begin{eqnarray}
p(\vec{r})=e^{\imath r\cos{\theta}}+\sum_{i=1}^N\sum_nA_n^j{\text{H}}_n(k|\vec{r}-\vec{r}_j|)e^{\imath n\theta_{(\vec{r}-\vec{r}_j)}}.\nonumber\\
\end{eqnarray}

\subsection{Scattering cross section}
From the previous equations, the expression of the scattering cross section of an array of scatterers can be obtained. The scattered pressure field by a distribution of scatterers can be written as
 \begin{eqnarray}
p_s(\vec{r})= \sum_{i=1}^N\sum_nA_n^j{\text{H}}_n(k|\vec{r}-\vec{r}_j|)e^{\imath n\theta_{(\vec{r}-\vec{r}_j)}}.
 \end{eqnarray}
 In the far field, we have
 \begin{eqnarray}
 {\text{H}}_n(k|\vec{r}-\vec{r}_j|) \simeq \sqrt{\frac{k}{\imath 2\pi |\vec{r}|}}(-\imath)^ne^{\imath k |\vec{r}|}e^{-\imath k |\vec{r_j}|\cos{(\theta-\theta_{\vec{r}_i})}},\nonumber\\
 \end{eqnarray}
 considering that $|\vec{r}-\vec{r}_j|\simeq |\vec{r}|-|\vec{r}_i|\cos{(\theta-\theta_{\vec{r}_i})}$. The far-field scattered pressure expression is also 
\begin{eqnarray}
p_{s}^{f}=S(\theta,\omega)\sqrt{\frac{k}{\imath 2\pi r}}e^{\imath k r},\;\;r\rightarrow\infty,
\end{eqnarray}
with the far-field scattered amplitude
\begin{eqnarray}
S(\theta,\omega)=\frac{2}{k} \sum_{i=1}^N e^{-\imath k |\vec{r}_i| \cos{(\theta-\theta_{\vec{r}_i})}} \sum_{n}(-i)^nA_n^ie^{\imath n\theta}.\nonumber\\
\end{eqnarray}
The scattering cross section is thus written as
\begin{eqnarray}
\sigma=-2\textrm{Re}(S(0,\omega))=-\frac{4}{k}\textrm{Re} \sum_{i=1}^N e^{-\imath k x_{\vec{r}_i}} \sum_{n}(-i)^nA_n^i.\nonumber\\\label{eq:sigma}
\end{eqnarray}

\begin{figure}
\includegraphics[width=85mm]{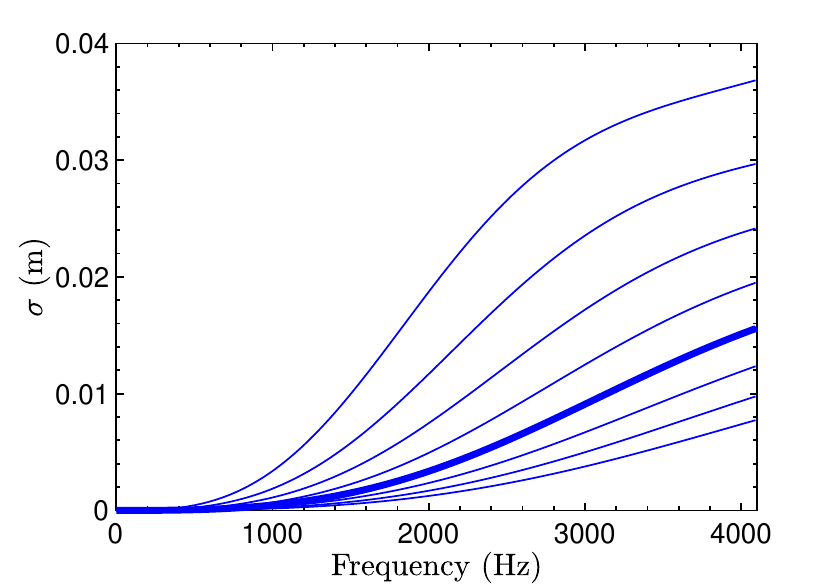}
\caption{Scattering cross sections for cylinders with radius [1/130, 1/120, 1/110, 1/100, 1/90, 1/80, 1/70, 1/60]. As the radius increases the scattering cross section increases also. The wide line corresponds to the scattering cross section of the cylinders analyzed in this work.}
\label{fig:scs}
\end{figure}

\subsection{Weak scattering approximation}
Strictly speaking the Born approximation implies that the interaction between the scatterers is negligible, in other words, the term $p_s(\vec{r},\vec{r}_j)=0$ in Eq. (\ref{eq:born1}), i.e., the incident wave on the $i$-th cylinder is only the incident wave without any contribution of the other scatterers in the structure. In this work, we will consider that $|p_s(\vec{r},\vec{r}_j)|<<|p_0(\vec{r})|$ $\forall j$. For a single scatterer, the scattering cross section, is defined as
\begin{eqnarray}
\displaystyle
\sigma=\frac{1}{{|p_0|^2}}\int|p_s|^2\textrm{d}s=\oint\frac{\textrm{d}\sigma}{\textrm{d}\Omega}\textrm{d}\Omega,
\end{eqnarray}
where the integral runs over a closed surface enclosing the scatterer. This could be used to evaluate the intensity of the scattered field by a single element. Figure \ref{fig:scs} shows the scattering cross section for scatterers with radius [1/130, 1/120, 1/110, 1/100, 1/90, 1/80, 1/70, 1/60]. We have chosen a configuration for which the scattering cross section is less than 0.015 in the range of frequencies analyzed in the work, meaning that the scattering is thus 1.5\% of the incident wave, so the weak scattering approximation is valid.

\section{Structure factor}
\label{app:SG}

Let us consider the scattering of an acoustic beam of wavelength $\lambda$ by the distribution of $N$ scatterers. We assume that the scattering is weak, so that the amplitude of the incident beam is higher than the amplitude of the scattering waves; absorption, refraction and higher order scattering can be neglected (kinematic diffraction). The direction of any scattered wave is defined by its scattering vector $\vec{G}=\vec{k_{s}} -\vec{k_{0}}$, where $\vec{k}_{s}$ and $\vec{k}_{0}=k_0(\cos{\theta^i},\sin{\theta^i})$ are the scattered and incident beam wavevectors with $\theta^i$ the incidence angle. For elastic scattering, $|\vec {k} _{s}|=|\vec {k}_{0} |=|\vec {k} |=2\pi /\lambda$ and then $G=|\vec {G} |={{\frac {4\pi }{\lambda }}\sin(\theta)}$. The amplitude and phase of this scattered wave is the vectorial sum of the scattered waves by all the scatterers $\Psi _{s}(\vec {q} )=\sum _{i=1}^{N}f_{i}{e} ^{-\imath\vec {G} \vec {r} _{i}}$, with $f_i$ the atomic structure factor. The scattered intensity reads as
\begin{eqnarray}
I(\vec {G} )&=&\Psi _{s}(\vec {G} ).\Psi _{s}^{*}(\vec {G} )\nonumber\\
&=&\sum _{j=1}^{N}f_{j} {e} ^{-i\vec {G} \vec {r} _{j}}\times \sum _{k=1}^{N}f_{k} {e} ^{i\vec {G} \vec {r} _{k}}\nonumber\\
&=&\sum _{j=1}^{N}\sum _{k=1}^{N}f_{j}f_{k} {e} ^{-\imath \vec {G} (\vec {r} _{j}-\vec {r} _{k})}.
\end{eqnarray}
The structure factor, $S(\vec{G})$, is then defined as this intensity normalized by $1/\sum _{j=1}^{N}f_{j}^{2}$
\begin{eqnarray}
S(\vec {G} )={\frac {1}{\displaystyle{\sum _{j=1}^{N}f_{j}^{2}}}}\sum _{j=1}^{N}\sum _{k=1}^{N}f_{j}f_{k}{e} ^{-\imath \vec {G} (\vec {r} _{j}-\vec {r} _{k})}.
\end{eqnarray}
If all the scatterers are identical, then
\begin{eqnarray}
I(\vec {G} )=f^{2}\sum _{j=1}^{N}\sum _{k=1}^{N}\ {e} ^{-\imath \vec {G} (\vec {r} _{j}-\vec {r} _{k})},
\end{eqnarray}
so
\begin{eqnarray}
S(\vec{G})={\frac {1}{N}}\sum _{j=1}^{N}\sum _{k=1}^{N} {e} ^{-\imath \vec{G} (\vec{r} _{j}-\vec{r} _{k})} = {\frac {1}{N}}\left|\sum _{j=1}^{N}{e} ^{\imath \vec{G} \vec{r} _{j}}\right|^2 .
\end{eqnarray}
Therefore, the structure factor $S(\vec{G})$ is proportional to the intensity of scattered field by a configuration of $N$ scatterers. It is worth noting here that the structure factor can be also related to the scattering cross section as follows
\begin{eqnarray}
\frac{d\sigma}{d\Omega}=f^{2}\sum _{j=1}^{N}\sum _{k=1}^{N}\ {e} ^{-\imath \vec {G} (\vec {r} _{j}-\vec {r} _{k})}=f^2NS(\vec{G}),
\end{eqnarray}
where $\sigma$ is the total cross-section and $\Omega$ is the solid angle.


Note that the von Laue condition\cite{kittel04, Ashcroft1976} for the periodic systems implies that the constructive interferences will occur if the difference between the incident and reflected wavevector is a vector of the reciprocal lattice. Therefore the Bragg scattering condition reads as $|\vec{k}|=\frac{|\vec{G}|}{2\sin{\theta}}$. In the 1D case ($\theta=\pi/2$), the wavevectors are collinear and then, $|\vec{k}|=|\vec{G}|/2$.

%

\end{document}